\RequirePackage[2020-02-02]{latexrelease}
\documentclass[preprint,aps,tightenlines,showkeys,nofootinbib,superscriptaddress]{revtex4}
\usepackage{latexsym}
\usepackage[dvips]{color}
\usepackage{graphicx}
\usepackage{makecell}
\newcommand{\beq}{\begin{eqnarray}}
\newcommand{\eeq}{\end{eqnarray}}
\newcommand{\bea}{\begin{eqnarray*}}
\newcommand{\eea}{\end{eqnarray*}}
\newcommand{\eq}{eqnarray}

\newcommand{\be}{{\beta}}

\newcommand{\ep}{{\epsilon}}

\newcommand{\de}{{\delta}}

\newcommand{\la}{{\lambda}}
\newcommand{\La}{{\Lambda}}

\newcommand{\om}{{\omega}}

\newcommand{\pa}{{\partial}}
\newcommand{\no}{{\nonumber}}
\newcommand{\f}{\frac}
\newcommand{\ra}{\rightarrow}

\newcommand{\Sch}{Schwarzschild }

\newcommand{\Ho}{Ho\v{r}ava}
\newcommand{\diff}{diffeomorphism}

\begin{document}

\preprint{arXiv:2607.xxxx [hep-th]}

\title{
Ho\v{r}ava Stars
Revisited:
New Phases of
Incompressible
Stars and Black Holes,
and
Buchdahl's Theorem}
\author{Mu-In Park\footnote{E-mail address: muinpark@gmail.com} 
}
\affiliation{
Center for Quantum Spacetime, Sogang University,
Seoul, 121-742, Korea}

\date{\today}

\begin{abstract}
I study a particular exact solution for {\it static}
stars in four-dimensional {\it non-projectable} Ho\v{r}ava gravity, which has been proposed
as a renormalizable
gravity model without the ghost problem by abandoning Einstein's equal-footing treatment
of space and time through anisotropic scaling with
$z>1$.
Considering the spherically symmetric static black-hole solutions in $z=3$ \Ho~gravity as the exterior spacetimes of stars, I obtain an exact solution for incompressible ({\it i.e.}, uniform-density) static stars with an arbitrary
cosmological constant and isotropic pressure, and $\la=1$, in which Birkhoff's theorem holds. For a vanishing cosmological constant, I obtain a modified
Buchdahl bound on the maximum compactness
for uniform-density stars, $4/9 \leq C_{\rm{max}} \leq 1$, ranging from $C_{\rm{max}}=4/9$
in the general relativity
limit to $C_{\rm{max}}=1$ in the {extremal-black-hole} limit.
These solutions have positive pressure but violate the dominant energy condition. By contrast,
I find that {\it Ultra-Compact Objects (UCOs)} with compactness $C>1$ also
exist with {\it negative pressure} while, surprisingly, satisfying {\it all} four standard energy conditions: null, weak, strong, and dominant energy conditions.
In addition to {\it mini stars} with masses below the extremal black-hole mass, UCOs include the
{\it regular (non-singular) black-hole} solutions with masses {above} the extremal black-hole mass.
In these solutions, the
matter is localized at the timelike core  region, $0<r<r_-$, bounded by
the inner horizon $r_-$, while their exterior metrics are unaffected by
the core matter and identical to the corresponding
{\it vacuum} \Ho~black-hole solutions. These long-sought regular black-hole solutions are essential
manifestations of Birkhoff's theorem in \Ho~gravity.
I also find {\it negative-mass} stars with {positive} pressure
that violate all standard energy conditions.
Finally, I prove
Buchdahl's
theorem in non-projectable Ho\v{r}ava gravity, which states that {\it the
compactness of any physically reasonable, static, spherically symmetric, and isotropic star is bounded above
by the modified Buchdahl bound
obtained for uniform-density Type-I stars}. The proof uses a weight-monotonicity condition on the average
density and Birkhoff's theorem, together with the usual assumption that average density
is non-increasing.

\end{abstract}

\keywords{Incompressible static \Ho~stars, regular \Ho~black holes, non-projectable \Ho~gravity, Buchdahl's bound, Buchdahl's theorem}

\maketitle

\newpage

\section{Introduction}

Ho\v{r}ava gravity has been proposed as a renormalizable
higher-derivative quantum gravity model without the ghost
problem, by abandoning the Lorentz symmetry from non-equal-footing treatment of space
and time through the anisotropic scaling with
 $z>1$ \cite{Hora:2009,Lifs:1941,DeWi:1967}. There have been
 numerous
 studies of its black hole solutions; in particular, several exact spherically symmetric black
 hole solutions were found \cite{Lu:2009,Keha:2009,Park:0905,Kiri:2009}, shortly after the appearance
 of Ref. \cite{Hora:2009}. Most recently, an {\it exact} rotating black hole solution has also been
 found in the low-energy sector of four dimensional non-projectable Ho\v{r}ava gravity
 \cite{Deve:2024}. Finding the exact rotating black hole solutions in the full
 four-dimensional (renormalizable) Ho\v{r}ava gravity is still an open question, although
 a {\it four}-dimensional {\it massless} rotating solution \cite{Park:2023} and {\it three}-dimensional rotating black hole solutions \cite{Park:2012,Soti:2014} have been found.

On the other hand, the corresponding ``star" solutions, whose gravitational collapse {\it could}
produce the \Ho~black hole solutions, have not yet been studied extensively. There are
{\it no-go} results for {\it static} stars in the {\it projectable} case with the space-independent lapse function $N=N(t)$ \cite{Izum:2009,Gree:2009}. In this paper, I consider the {\it non-projectable} case with the generic space- and time-dependent lapse $N=N(t,{\bf x})$, in a closer parallel with general relativity (GR) and with the exact Ho\v{r}ava black hole solutions in Refs. \cite{Lu:2009,Keha:2009,Park:0905,Kiri:2009}.

The organization of the paper is as follows. In Sec. II, I consider four-dimensional $z=3$
non-projectable
\Ho~gravity coupled to matter and obtain the reduced action and the equations of motion for a static, spherically symmetric metric ansatz with perfect-fluid matter. In Sec. III, I consider the $\la=1$ case, in which Birkhoff's theorem holds, and derive a Tolman-Oppenheimer-Volkoff-like equation of hydrostatic equilibrium. Assuming isotropic pressure, I obtain an exact solution for incompressible (uniform-density) stars with an arbitrary cosmological constant. In Sec. IV, I consider the case of a vanishing cosmological constant and obtain the modified Buchdahl bound on the maximum compactness of uniform-density Type-I stars. I also find additional stellar phases--Type-II and III stars--as well as regular (non-singular) \Ho~black-hole solutions, for which the Buchdahl bound does not exist but other new physical bounds do. The new phases include ultracompact objects with compactness $C>1$ among Type-II stars, as well as regular black-hole solutions. In Sec. V, I prove Buchdahl's theorem in \Ho~gravity for Type-I stars using a weight-monotonicity condition on the average density and Birkhoff's theorem, together with the usual assumption that the average density is non-increasing. In Sec. VI, I conclude with some remarks on open problems. In Appendix {\bf A}, I present details of the energy conditions.

\section{Non-projectable Ho\v{r}ava gravity coupled to  matter}

Considering the ADM
metric
\begin{\eq}
\label{metric}
ds^2=-N^2 c^2 dt^2+g_{ij}\left(dx^i+N^i dt\right)\left(dx^j+N^j
dt\right),\
\end{\eq}
the non-projectable $z=3$ Ho\v{r}ava gravity coupled to matter is described by the action \cite{Hora:2009,Kiri:2009b,Char:2009}, up to boundary terms,
\begin{\eq}
S &= & \int dt d^3 x
\sqrt{g}N\left[\frac{2}{\kappa^2}\left(K_{ij}K^{ij}-\lambda
K^2\right)-\frac{\kappa^2}{2\nu^4}C_{ij}C^{ij}+\frac{\kappa^2
\mu}{2\nu^2}\epsilon^{ijk} R^{}_{i\ell} \nabla_{j}R^{\ell}{}_k
\right.
\nonumber \\
&&\left. -\frac{\kappa^2\mu^2}{8} R^{}_{ij}
R^{ij}+\frac{\kappa^2 \mu^2}{8(3\lambda-1)}
\left(\frac{4\lambda-1}{4}R^2-\Lambda_W R^{}+3
\Lambda_W^2\right)+\frac{\kappa^2 \mu^2 \om}{8(3\lambda-1)}
R^{}\right]\ +S_M, \label{horava}
\end{\eq}
where
$
 K_{ij}=(2N)^{-1}(\dot{g}_{ij}-\nabla_i N_j-\nabla_jN_i)
$
is the extrinsic curvature,
$
 C^{ij}=\epsilon^{ik\ell}\nabla_k
(R^{j}{}_\ell-R^{} \delta^j_\ell/4)
$
is the Cotton tensor, and $\kappa,\lambda,\nu,\mu$, and $\La_W$
are coupling constants,
and $\om$ is an IR-modification parameter that softly breaks
the detailed-balance condition so that Newtonian gravity or the GR limit exists in the IR, without changing the improved UV behavior \cite{Hora:2009,Keha:2009,Park:0905}. Here, $\epsilon^{ikl}$ is the Levi-Civita symbol, the overdot $(\dot{})$ denotes the time derivative, $\nabla_i$ is the covariant derivative for the induced metric $g_{ij}$ on the time-slicing hypersurface $\Sigma_t$, and $R_{ij}$ and $R$ are the three-dimensional (Euclidean) Ricci tensor and scalar, respectively.

Here, the Lorentz violation is introduced by the IR Lorentz-deformation parameter $\la$ $(\la \neq 1/3)$
and by non-covariant deformations with higher-order spatial derivatives (up to sixth order), yielding
a renormalizable gravity theory without the ghost problem in the UV while still recovering GR in the IR under appropriate limits \cite{Argu:2015}.

The equations obtained by varying the {generic (non-projectable)} $N$ and $N^i$ are given by
\beq
-\f{1}{\sqrt{g}}\f{\de S_M}{\de N}&=&\f{2}{\kappa^2}(K_{ij}K^{ij} -\lambda K^2) -\frac{\kappa^2\mu^2
\left[(\Lambda_W - \omega) R -3\Lambda_W^2 \right]}{8(1-3\lambda)}-\frac{\kappa^2\mu^2 (1-4\lambda)}{32(1-3\lambda)}R^2 \no \\
&&+ \frac{\kappa^2}{2\nu^4} Z_{ij} Z^{ij}\, , \label{eom1} \\
-\f{1}{\sqrt{g}}\f{\de S_M}{\de N_i}&=&\f{2}{\kappa^2}\nabla_k(K^{ki}-\lambda\,Kg^{ki})\, ,\label{eom2}
\eeq
which correspond to the Hamiltonian and momentum constraints, respectively, where
$
Z_{ij}\equiv C_{ij} - {\mu \nu^2} R_{ij}/{2}.
$
The equations of motion obtained by varying $g^{ij}$ are quite messy and are given by \cite{Lu:2009,Kiri:2009b,Char:2009,Deve:2018}:
\beq
-\f{1}{\sqrt{g}}\f{\de S_M}{\de g_{ij}}
&=&\frac{2}{\kappa^2}E_{ij}^{(1)}-\frac{2\lambda}{\kappa^2}E_{ij}^{(2)}
+\frac{\kappa^2\mu^2(\Lambda_W-\omega)}{8(1-3\lambda)}E_{ij}^{(3)}
+\frac{\kappa^2\mu^2(1-4\lambda)}{32(1-3\lambda)}E_{ij}^{(4)} \no \\
&&-\frac{\mu\kappa^2}{4\nu^2}E_{ij}^{(5)}
-\frac{\kappa^2}{2\nu^4}E_{ij}^{(6)} \no \\
&\equiv&\frac{2}{\kappa^2}E_{ij},
\label{eom3}
\eeq
where
\bea
E_{ij}^{(1)}&=& N_i \nabla_k K^k{}_j + N_j\nabla_k K^k{}_i -K^k{}_i
\nabla_j N_k-
   K^k{}_j\nabla_i N_k - N^k\nabla_k K_{ij}\no\\
&& - 2N K_{ik} K_j{}^k
  -\frac{1}{2} N K^{k\ell} K_{k\ell}\, g_{ij} + N K K_{ij} + \dot K_{ij}
\,,\no \\
E_{ij}^{(2)}&=& \frac{1}{2} NK^2 g_{ij}+ N_i \pa_j K+
N_j \pa_i K- N^k (\pa_k K)g_{ij}+  \dot K\, g_{ij}\,,\no\\
E_{ij}^{(3)}&=&N\left(R_{ij}- \frac{1}{2}R g_{ij}+\frac{3}{2}
\frac{\Lambda_W^2}{\Lambda_W-\omega} g_{ij}\right)-(
\nabla_i\nabla_j-g_{ij}\nabla_k\nabla^k)N\,,\no\\
E_{ij}^{(4)}&=&NR\Big(2 R_{ij}-\frac{1}{2}R g_{ij}\Big)- 2
\big(\nabla_i\nabla_j-g_{ij}\nabla_k\nabla^k\big)(NR)\,,\no\\
E_{ij}^{(5)}&=&\nabla_k\big[\nabla_j(N Z^k_{~~i}) +\nabla_i(N
Z^k_{~~j})\big]  -\nabla_k\nabla^k(N Z_{ij})
-\nabla_k\nabla_\ell(N Z^{k\ell})g_{ij}\,, \no \\
E_{ij}^{(6)}&=&-\frac{1}{2}N Z_{k\ell}Z^{k\ell}g_{ij}+
2NZ_{ik}Z_j^{~k}-N(Z_{ik}C_j^{~k}+Z_{jk}C_i^{~k})
+NZ_{k\ell}C^{k\ell}g_{ij}\no\\
&&-\frac{1}{2}\nabla_k\big[N\epsilon^{mk\ell}
(Z_{mi}R_{j\ell}+Z_{mj}R_{i\ell})\big]
+\frac{1}{2}R^n{}_\ell\, \nabla_n\big[N\epsilon^{mk\ell}(Z_{mi}g_{kj}
+Z_{mj}g_{ki})\big]\no\\
&&-\frac{1}{2}\nabla_n\big[NZ_m^{~n}\epsilon^{mk\ell}
(g_{ki}R_{j\ell}+g_{kj}R_{i\ell})\big]
-\frac{1}{2}\nabla_n\nabla^n\nabla_k\big[N\epsilon^{mk\ell}
(Z_{mi}g_{j\ell}+Z_{mj}g_{i\ell})\big]\no\\
&&+\frac{1}{2}\nabla_n\big[\nabla_i\nabla_k(NZ_m^{~n}\epsilon^{mk\ell})
g_{j\ell}+\nabla_j\nabla_k(NZ_m^{~n}\epsilon^{mk\ell})
g_{i\ell}\big]\no\\
&&+\frac{1}{2}\nabla_\ell\big[\nabla_i\nabla_k(NZ_{mj}
\epsilon^{mk\ell})+\nabla_j\nabla_k(NZ_{mi}
\epsilon^{mk\ell})\big]
-\nabla_n\nabla_\ell\nabla_k
(NZ_m^{~n}\epsilon^{mk\ell})g_{ij}\,. \\
\eea

Let me consider a static, spherically symmetric metric ansatz (hereafter, I shall use units with $c=1$ unless stated otherwise):
\begin{\eq}
  ds^2=-N(r)^2 dt^2+\frac{dr^2}{f(r)}+r^2
\left(d\theta^2+\sin^2\theta d\phi^2\right)\ \label{ansatz}.
\end{\eq}
Then, after substituting the metric ansatz into the action (\ref{horava}) and performing the angular integration, the
reduced action becomes
\begin{\eq}
{S_{red}}&=&\frac{\kappa^2\mu^2 \Omega }{8(1-3\lambda)}\int dt dr \frac{N}{\sqrt{f}}\left[(2\lambda-1)\frac{(f-1)^2}{r^2}
-2\lambda\frac{f-1}{r}f'+\frac{\lambda-1}{2}f'^2\right.\nonumber
\\ &&\left. ~~-2 (\om-\La_W) (1-f-rf') - 3 \La_W^2 r^2 \right]+S_M\ ,
\end{\eq}
where the prime $(')$ denotes differentiation with respect to $r$, and $\Omega=4 \pi$ is the solid angle.

The equations of motion obtained by varying $N$ and $f$ are
\begin{\eq}
\rho &=&\frac{\kappa^2 \mu^2}{8(1-3\lambda)} \f{1}{r^2} \left[(2\lambda-1)\frac{(f-1)^2}{r^2}-
2\lambda\frac{f-1}{r}f'+\frac{\lambda-1}{2}f'^2  \right. \no \\
&&\left. -2 (\om-\La_W) (1-f-r f')- 3 \La_W^2 r^2 \right]\ \label{Eq_rho},
\\
\f{N r^2}{2 f^{3/2}} p_r &=& -\frac{\kappa^2 \mu^2}{8(1-3\lambda)} \f{N}{r^2 f^{3/2}} \left\{\left(\frac{N'}{N}\right)
f \left[(\lambda-1) r^2 f'-2\lambda r(f-1)+2(\om-\La_W)r^3\right] \right. \no \\
&&+\f{1}{4} \left[ 2 (2 \la-1) (f-1)^2 - (\lambda-1) r^2 f'^2-4 (\om-\La_W)r^2 (1-f)-6 \La_W^2 r^4\right] \no \\
&&\left.+(\lambda-1)f\left[r^2 f''-2(f-1)\right]\right\}, \label{Eq_p}
\end{\eq}
respectively \footnote{Eqs. (\ref{Eq_rho}) and (\ref{Eq_p}) imply that $f$ and $N$ are at least $C^1$ as long as $\rho$ and $p_r$ are continuous. For arbitrary $\la \neq 1$, Eq. (\ref{Eq_p}) shows that
a discontinuity in $\rho$ at a certain $r$ produces a delta-function singularity in $N'$ unless
$p_r$ is also singular there. On the other hand, only for $\la=1$, $f$ is $C^0$ and $N$ is $C^1$
when $\rho$ is discontinuous but $p_r$ is continuous. }. In addition,
the fluid satisfies the continuity equation
\beq
p'_r=-N^{-1} N' (p_r+\rho)-\f{1}{r} \left(2 p_r -p_{\theta}-p_{\phi} \right)
\label{P'}
\eeq
from the {\it spatial} component of the four-dimensional energy-momentum conservation law $\hat{\nabla}_{\mu} T^{\mu i}=0$ (where $\hat{\nabla}_{\mu}$ is the four-dimensional covariant derivative) \cite{Kiri:2009b,Char:2009}. This law is still valid
even though the action (\ref{horava}) lacks
full diffeomorphism (${\it Diff}$) invariance and its manifest symmetry is the {\it foliation-preserving} \diff~(${\it Diff}_{\cal F}$) \cite{Hora:2009}.
Here, I consider perfect-fluid matter in
spherically symmetric metric (\ref{ansatz}) \footnote{The general formula for  non-vanishing shift $N^i$ will be $T_{0i}=-\f{N}{\sqrt{g}}\f{\de S_M}{\de N^i}+T_{ij}N^j, T_{00}=-\f{N^2}{\sqrt{g}}\f{\de S_M}{\de N}+2 T_{0i}N^i-T_{ij} N^i N^j$ with $T_{ij}$ in Eq. (\ref{Tij}) \cite{Gour:2007}.} with the energy-momentum tensor
\beq
&&T_{00}=-\f{N^2}{\sqrt{g}}\f{\de S_M}{\de N}=N^2 \rho,~T_{0i}=-\f{N}{\sqrt{g}}\f{\de S_M}{\de N^i}=0, \no \\
&&T_{ij}=-\f{2}{N\sqrt{g}}\f{\de S_M}{\de g^{ij}}=p_i \delta_{ij}.\label{Tij}
\eeq

\section{Tolman-Oppenheimer-Volkoff-like equation and incompressible Ho\v{r}ava star solutions}

In order to study explicit star solutions, let me consider the $\la=1$ case, for which
the Ho\v{r}ava gravity action (\ref{horava}) and its
static black hole solutions reduce to GR and the \Sch solution, respectively, in the IR
limit \cite{Argu:2015}, and
Birkhoff's theorem is
valid \cite{Deve:2018}. Then, inspired by the
exact black hole solutions in Refs. \cite{Lu:2009,Keha:2009,Park:0905}, I posit the
following static interior stellar solution for $f$:
\beq
f(r)=1+(\om-\La_W) r^2-\sqrt{r[\omega (\om-2 \La_W) r^3 + \be (r)]},
\label{f(r)}
\eeq
then, by solving Eq. (\ref{Eq_rho}), I obtain
\beq
\be (r) &=&\f{16}{\kappa^2 \mu^2} \int_0^r \rho (r') r'^2 dr' \no \\
&\equiv &
4 \om m (r),
\label{beta_sol}
\eeq
where $m(r)$ is the star's mass function
and $r$ is space-like, {\it i.e.}, $f(r)>0$, from the {\it staticity} condition, in closely parallel with the GR case \cite{Wald:1984}.

For $\la=1$, Eq. (\ref{Eq_p}) can be written as
\beq
\frac{N'}{N}=\f{-\frac{4}{ \kappa^2 \mu^2} r^3 p_r +\f{r}{2}(\om-\La_W) (f-1)
+\f{1}{4 r}(f-1)^2 -\f{3}{4} \La_W^2 r^3}{f \left[ f-1-(\om-\La_W) r^2 \right]}.\label{N'}
\eeq
Then, the continuity equation (\ref{P'}) becomes
\beq
p'_r&=&-\f{(p_r+\rho)\left[-\frac{4}{ \kappa^2 \mu^2} r^3 p_r +\f{r}{2}(\om-\La_W) (f-1)
+\f{1}{4 r}(f-1)^2 -\f{3}{4} \La_W^2 r^3 \right]}{f \left[ f-1-(\om-\La_W) r^2 \right]}
\label{TOV2}
\\
&&-\f{1}{r} \left(2 p_r -p_{\theta}-p_{\phi} \right), \no
\eeq
which is a Tolman-Oppenheimer-Volkoff (TOV)-like equation of hydrostatic equilibrium \cite{Tolm:1939,Oppe:1939}.
Here, it is important to note that, because of our choice of $\la=1$, Eqs. (\ref{N'})
and (\ref{TOV2}) are algebraic in $\rho(r), p_i(r)$, and $f(r)$ (or $m(r)$). Thus,
even with the higher-derivative contribution, `` $(f-1)^2$ '',
the system still consists of two {\it first}-order ODEs, as in GR. This is the key property that makes
the system easier to integrate analytically or numerically \cite{Park:2010,Kim:2018},
in contrast to other (covariant) higher-curvature
theories: if I had chosen $\la \neq 1$, I would have the additional dependence on $f'(r), f''(r)$ or $m''(r)$ and the system would become a set of {\it higher}-order ODEs, which would be much more complicated to analyze, as in other higher-curvature modified theories of gravity \cite{Pani:2011}.

Given the importance of isotropic incompressible (or uniform-density) stars in GR, which leads the maximum mass limit for static stars of a fixed radius, I now seek incompressible Ho\v{r}ava star solutions with uniform density $\rho_0$ and radius $R$
\beq
\rho(r)= \left\{\begin{array}{ll}
\rho_0 & {(r \le R)} \\
0  & {(r > R)}
\end{array}\right.
\eeq
and hence
\beq
\be(r)=\f{16}{3 \kappa^2 \mu^2} \rho_0 r^3~~(r \le R)
\eeq
with the  isotropic pressure, $p_r=p_{\theta}=p_{\phi}\equiv p$. Then, the TOV-like equation (\ref{TOV2}) can be written as
\beq
p'=-\f{(p+\rho)r \left[-\frac{4}{ \kappa^2 \mu^2} p +\f{1}{4}(Q+\La_W)(Q-3\La_W)
+\f{1}{2}Q \omega \right]}{( Q-\om+\La_W)(1+Q r^2 )}, \label{P'_uniform}
\eeq
where I used
\beq
Q \equiv \om-\La_W-\sqrt{\omega (\om-2 \La_W) +\f{16}{3 \kappa^2 \mu^2} \rho_0 }
\eeq
and
\beq
f(r)=1+Q r^2 \label{f(Q)}.
\eeq
Eq. (\ref{P'_uniform}) can be integrated to obtain the solution,
\beq
p(r)=\rho_0 \left[\f{\sqrt{1+Q R^2}-\sqrt{1+Q r^2}}{\sqrt{1+Q r^2}-\rho_0 \left(\f{a}{b}\right) \sqrt{1+Q R^2}} \right], \label{P_sol_AdS}
\eeq
where the integration constant $\sqrt{1+Q R^2}$ was fixed by the boundary condition $p(R)=0$,
at the stellar surface $r=R$; the constant parameters are
\beq
 a \equiv -\f{4}{\kappa^2 \mu^2},~ b \equiv \f{1}{4}(Q+\La_W)(Q-3\La_W)
+\f{1}{2}Q \omega.
\eeq

Substituting the solution (\ref{P_sol_AdS}) into Eq. (\ref{N'}), I obtain
\beq
N(r)=\f{\sqrt{1+Q r^2}-\rho_0 \left(\f{a}{b}\right) \sqrt{1+Q R^2}}{1-\rho_0 \left(\f{a}{b}\right)}  \label{N_sol_AdS}
\eeq
by matching to
the exterior vacuum solution \cite{Lu:2009,Keha:2009,Park:0905}
\beq
N^2_{ext} (r)=f_{ext}(r)=1+(\om-\La_W) r^2-\sqrt{r[\omega (\om-2 \La_W) r^3 + 4 \om M]}
\label{f(R)}
\eeq
at $r=R$, where the
star's mass is $M=4 \pi R^3 \rho_0/3$. It is rather remarkable that the structures of the solutions (\ref{P_sol_AdS}) and (\ref{N_sol_AdS}) are almost the same as in GR, despite the quite different forms of the interior stellar solution (\ref{f(r)}) and the exterior vacuum solution (\ref{f(R)}) due to the Lorentz-violating higher-derivative terms in the Ho\v{r}ava gravity action (\ref{horava}). Moreover, the existence of the {\it static} isotropic solutions is in sharp contrast to the {\it no-go} result in the {\it projectable} case \cite{Izum:2009,Gree:2009}.

Finally, I note that, due to the {\it effectively (A)dS} structure of the interior stellar
metric (\ref{f(Q)}), there are no curvature singularities, as shown by the three-curvature invariants as
\beq
R(r)=-6Q,~~ R^{ij}R_{ij}(r)=12 Q^2,
\label{R}
\eeq
which are finite constants \footnote{The curvature invariants are discontinuous at $R$ since $f$ is $C^0$ (see footnote 1).} for any finite radius $R$. This contrasts with both the usual point-like singularity at $r=0$ and the unusual surface-like singularity at $r_S=(-4M/(\om-2 \La_W))^{1/3}$, if it exists, in the exterior vacuum solution (\ref{f(R)}) \cite{Argu:2015}.

\section{The Buchdahl's maximum-compactness bound, and
new phases of \Ho~stars and black holes
in asymptotically {flat}
spacetime}

In GR, incompressible
stars \cite{Schw:1916}, although their sound speeds are infinite,
$c^2_s=dp/d \rho \ra \infty$, and therefore violate the maximum speed limit, play an
important role because they
saturate the maximum-compactness bound (without a cosmological constant),
\beq
C\equiv M/R
 \leq 4/9,
\label{Buchdahl_GR}
\eeq
due to the maximum possible mass of a uniform-density star, $M=(4/9) R$, for which the central
pressure $p(0)$ in hydrostatic equilibrium becomes
infinite; this is known as the Buchdahl limit \cite{Buch:1959} \footnote{The uniform-density star solution and its maximum possible mass $M=(4/9) R$ were first obtained by Schwarzschild in 1916 \cite{Schw:1916}, but the result is nowadays called the Buchdahl limit after Buchdahl's 1959 proof of the bound (\ref{Buchdahl_GR}) for {\it any} static, isotropic stars under reasonable assumptions \cite{Buch:1959}.
In 1938 \cite{Zwik:1938}, Zwicky proposed collapsed neutron stars using the Schwarzschild's uniform-density stellar solution.}. However, in \Ho~gravity, the incompressible stars studied in the previous section, are still viable because there is no corresponding speed limit!

From the pressure solution (\ref{P_sol_AdS}) for the uniform-density star, the corresponding Buchdahl limit, at which the cental pressure becomes infinite, is given by
\beq
b=\rho_0 a \sqrt{1+Q R^2}
\label{Buchdahl_limit_0}
\eeq
provided that $\rho_0 ({a}/{b})>0$,
from the vanishing of the denominator of Eq. (\ref{P_sol_AdS}) at $r=0$.

For
asymptotically {\it flat} spacetime with $\La_W=0$, the usual convention $(\kappa^2 \mu^2)^{-1} \equiv \pi \omega G_N/c^4 >0$ \cite{Keha:2009}, and positive mass $M>0$, I have (hereafter, I use units $c=G_N=1$)
\beq
 a &\equiv&-4 \pi \om <0, \no \\
 Q &\equiv&\omega \left(1-\sqrt{1+\f{4M}{\om R^3}} \right) \leq 0,
\eeq
and hence the Buchdahl limit exists only when
\beq
b&\equiv&\f{1}{4}Q(Q+2\om) \no \\
 &=&\f{\om^2}{4} \left(1-\sqrt{1+\f{4M}{\om R^3}} \right)  \left(3-\sqrt{1+\f{4M}{\om R^3}} \right)<0,
 \label{b<0}
\eeq
{\it i.e.},
\beq
\om > \f{M}{2  R^3}
\label{b<0:omega}
\eeq
is satisfied.
Then, Eq. (\ref{Buchdahl_limit_0}) can be written as
\beq
-\f{1}{4}Q(Q+2\om)=\f{3 M \om}{R^3} \sqrt{1+Q R^2}
\label{Buchdahl_limit_Q}
\eeq
and I finally obtain an algebraic equation for the Buchdahl limit,
\beq
\sqrt{1+4 y}-1-y= 3y \sqrt{1+\om R^2 (1-\sqrt{1+4 y} )}
\label{Buchdahl_limit_Q}
\eeq
or
\beq
y^2 \left( a_8 y^3+a_6 y^2 +a_4y+a_2\right)=0,
\label{Buchdahl_limit_y}
\eeq
where $a_8=324 \om^2 R^4, a_6=-(64+288 \om R^2 ), a_4=(112-72 \om  R^2)$, and $a_2=32$, with the
condition $y \equiv {M}_{\rm{max}} (\om, R)/{\om R^3}<2$ from Eq. (\ref{b<0:omega}), or equivalently
$b<0$ in Eq. (\ref{b<0}); ${M}_{\rm{max}} (\om, R)$
is the maximum allowed mass for fixed $\omega >0$ and $R$.

The three roots of Eq. (\ref{Buchdahl_limit_y}), other than the trivial root $y=0$, are given by the cubic
formula \cite{Mann:2008}
\beq
y_0&=&s_{+}+s_{-}-\f{a_6}{3 a_8}, \no \\
y_{\pm}&=&-\f{1}{2} (s_{+}+s_{-})-\f{a_6}{3 a_8} \pm \f{\sqrt{3}}{2} (s_{+}-s_{-}),
\label{y_naive}
\eeq
where
\beq
s_{\pm}&=&3 \sqrt{-z \mp \sqrt{-q^2+z^2}}, \\
q&=&\f{-3 a_8 a_4+a_6^2}{9 a_8^2}, ~z=\f{9 a_8 a_6 a_4-27 a_8^2 a_2-2a_6^3}{54 a_8^3}. \no
\eeq

\begin{figure}
\includegraphics[width=11cm,keepaspectratio]{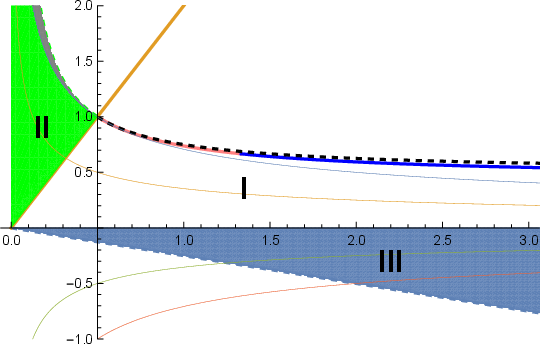}
\caption{The plot of the 
compactness
$M/R$ vs. $\om R^2$ for uniform-density stars
The thick red and blue curves denote the Buchdahl-limit branches $y_+$ and $y_0$ for $1/2 \leq \om R^2 \leq 4/3$ and $4/3 \leq \om R^2 < \infty$, respectively, in Eq. (\ref{y_true}) and give the maximal compactness bound
$4/9 \leq (M/R)_{\rm{max}} \leq 1$, with $(M/R)_{\rm{max}}=1$ in the extremal-black-hole limit $\om R^2\ra1/2$, and $(M/R)_{\rm{max}}= 4/9$ in the GR limit
$\om R^2\ra \infty$.
The dashed green and black curves denote inner and outer black hole horizons, respectively,
which merge at the extremal black hole $\om R^2=1/2$. The thick yellow line denotes
 $M/\om R^3=1/2$ which meets the black hole horizons and the Buchdahl limit curve
 simultaneously at the extremal black hole.
  The blue dashed line denotes the lower bound of the {\it real-valued} pressure for $M<0$. The (upper and lower) thin curves denote $M=constant$ for $M>0$ and $M<0$, respectively.}
\label{fig:C_vs_x}
\end{figure}

I note that the branches in Eq. (\ref{y_naive}), although real-valued, are  not all true roots of Eq. (\ref{Buchdahl_limit_Q}) but include {\it spurious} roots because Eq. (\ref{Buchdahl_limit_Q}) was squared twice to obtain Eq. (\ref{Buchdahl_limit_y}). The true root branches for the Buchdahl limit are found by the careful analysis of substituting the roots back into the original Eq.
(\ref{Buchdahl_limit_Q})
\footnote{In my earlier (unpublished) work \cite{Park:2010}, the spurious root branches were
overlooked, and an incorrect Buchdahl limit for $1/2<\om R^2<4/3$ was obtained. This was
recently corrected
by E. J. Son \cite{Son:2026}.
 }:
\beq
y
= \left\{\begin{array}{ll}
y_{+} & {(1/2 \leq \om R^2 \leq 4/3)} \\
y_{0} & {(4/3 \leq \om R^2 <\infty)}.
\end{array}\right.
\label{y_true}
\eeq
The resulting true branch is smooth at the crossover $\om R^2=4/3$ between $y_+$ and $y_0$
(the thick red and blue curves
in Fig. \ref{fig:C_vs_x}, respectively), but no true root branch exists for $\om R^2<1/2$ because the
condition $b<0$ in Eq. (\ref{b<0}), equivalently Eq. (\ref{b<0:omega}), is not satisfied.

The analytic form of the true root  branch for the Buchdahl limit (\ref{y_true}) is still complicated and difficult to express simply. However, if I write Eq. (\ref{Buchdahl_limit_y}) as
\beq
\left.\f{4 C}{x^4} \left[2 \left(4-9 C\right)x^2+ C
\left(28-72 C+81 C^2 \right) x-16 C^2\right]\right|_{C=C_{\rm{max}}}=0,
\eeq
where $C_{\rm{max}}(\om, R)=M_{\rm{max}}(\om, R)/R$ denotes the {\it maximal} compactness, the roots for $x\equiv \om R^2$ are given by
\beq
x_{\pm}=\f{-C_{\rm{max}} (28-72 C_{\rm{max}}+81 C_{\rm{max}}^2) \pm \sqrt{C_{\rm{max}}^2 (2-3 C_{\rm{max}})^2 (4-4 C_{\rm{max}}+9 C_{\rm{max}}^2)}}{4(4-9 C_{\rm{max}})},
\eeq
other than the trivial solution $x=\pm \infty$. In this case also, not all branches are true roots, and the spurious root branches need to be removed; the resulting true root branch can be expressed more simply as
\beq
x=\f{-C_{\rm{max}} (28-72 C_{\rm{max}}+81 C_{\rm{max}}^2) - 9 C_{\rm{max}} (2-3 C_{\rm{max}}) \sqrt{4-4 C_{\rm{max}}+9 C_{\rm{max}}^2}}{4(4-9 C_{\rm{max}})}
\label{x_true}
\eeq
when $4/9 \leq C_{\rm{max}} \leq 1$; otherwise, there are no true root branches for
the Buchdahl limit. The solution (\ref{x_true}) corresponds to the compactness curve for
the Buchdahl limit in Fig. \ref{fig:C_vs_x} in a simpler but {\it implicit} algebraic form.

For large $x=\om R^2 \gg 1$, using either $y_0$ in Eq. (\ref{y_true}) or Eq. (\ref{x_true}), the maximal compactness can be expanded as
\beq
C_{\rm{max}}(x)=\f{4}{9}+\left(\f{8}{27} \right) x^{-1}+ \left(\f{16}{729} \right) x^{-2}
+{\cal O} (x^{-3}),
\label{C_max_expansion}
\eeq
which reduces to the usual Buchdahl bound $C_{\rm{max}}=({4}/{9})$
in the GR limit $x=\om R^2 \ra \infty$. However, as shown in Fig. \ref{fig:C_vs_x}, for any finite $x >0$, $C_{\rm{max}}$ is {\it always greater} than its GR-limit value, {\it i.e.,} $C_{\rm{max}}>4/9$, whenever it exists. This means the uniform-density static stars in \Ho~gravity are more compact than those in GR. This is an expected result because more mass is required, at fixed radius, to form stars than in GR, owing to the weaker gravitational attraction caused by higher-derivative effects. This is a robust prediction of Ho\v{r}ava gravity that could be tested in the near future.

\begin{figure}
\includegraphics[width=5cm,keepaspectratio]{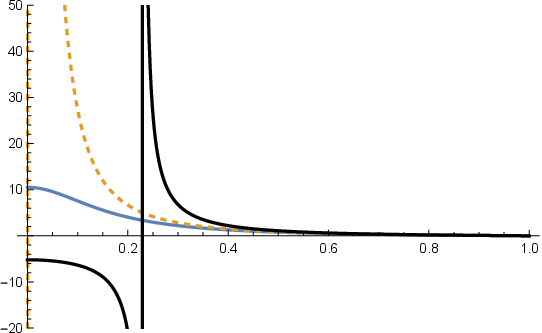}
\includegraphics[width=5cm,keepaspectratio]{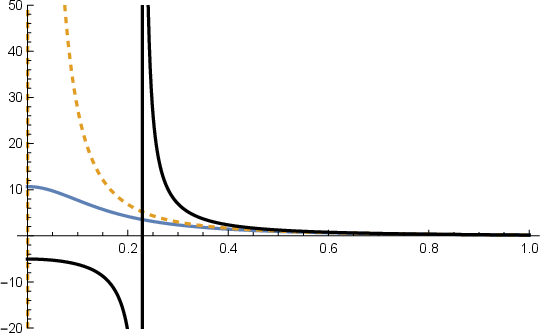}
\includegraphics[width=5cm,keepaspectratio]{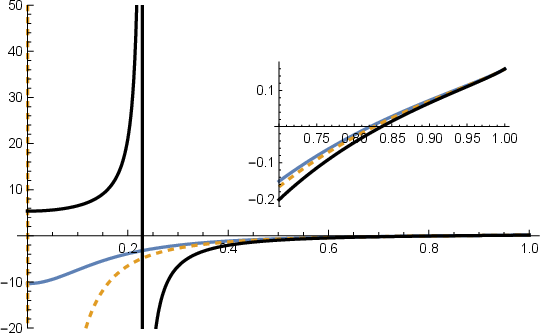}
\caption{The plots of the pressure $p(r)$ (left),
$\rho+p(r)$ (center), and $\rho-p(r)$ (right)
for the uniform-density stars with $M/\om R^3 <1/2$ or $b<0$ in asymptotically flat space-time with $\La_W=0$. The blue solid, yellow dashed, and black solid curves represent, respectively, $C (\equiv M/R) < C_{\rm{max}} (x),~ C = C_{\rm{max}} (x)$, and $C > C_{\rm{max}} (x)$. In the plots, we consider $C=5.995/9,~ 6/9,~ 6.01/9$, respectively, with $x \equiv \om R^2 =4/3$ and $C_{\rm{max}} (x)=6/9$.}
\label{fig:C>C_max}
\end{figure}

The {\it staticity} condition for stars requires $C_{\rm{max}}$ not to exceed the compactness of
black holes \cite{Keha:2009}
\beq
C_{bh}=\f{M_{bh}}{r_{\pm}}=\f{1}{2}+\f{1}{4 \om r_{\pm}^2}
\label{C_horizon}
\eeq
for the black hole's outer and inner horizons $r_{\pm}$ and its
mass $M_{bh}$; thus $C_{\rm{max}}<C_{\rm{bh}}$, as shown in Fig. \ref{fig:C_vs_x}.
In other words, the uniform-density static stars beyond the Buchdahl limit  $C_{\rm{max}}$ do not exist and there is a compactness gap between $C_{\rm{max}}$ and $C_{\rm{bh}}$. In the Buchdahl limit, the central pressure
$p(0)$ becomes infinite (the yellow dashed curve in Fig. \ref{fig:C>C_max}); beyond the
Buchdahl limit,
{\it i.e.}, $C_{\rm{max}}< C~ (<C_{\rm{bh}})$, the pressure $p(r)$ becomes infinite at another radius $r_*$, where the denominator of the pressure in Eq. (\ref{P_sol_AdS}) vanishes, $\sqrt{1+Q {r_*}^2}-\rho_0 \left(\f{a}{b}\right) \sqrt{1+Q R^2}=0$, while in its interior region $r<r_*$, the pressure $p(r)$ becomes {\it negative}, with $\rho+p<0$, violating the null energy condition (NEC) (the black solid curve in Fig. \ref{fig:C>C_max}).

However, as $\om R^2$ approaches to $1/2$,
$C_{bh}$ for black holes and $C_{\rm{max}}$ for uniform-density stars
merge and meet at
$(M_{\rm{max}}/R)=(M_{bh}/{r_{\pm}})=1$, where the black hole becomes extremal and its outer and inner horizons coincide, $r_+=r_-$. In otherwords,
the compactness gap between the Ho\v{r}ava black holes and stars vanishes as the stars approach
extremal black holes, implying a {\it continuous} transition from uniform-density static
stars to extremal black holes  \cite{Son:2026,Son:2026b} ! \footnote{This result was not obtained in my earlier work \cite{Park:2010}, due to an
incorrect Buchdahl limit for $\om R^2 <4/3$ (see footnote 3).}

Furthermore, the line $b=0$, which corresponds to the critical boundary for the existence of the Buchdahl limit (\ref{b<0}) or (\ref{b<0:omega}), intersects both $C_{\rm{max}}$ and $C_{bh}$ simultaneously at the extremal black hole limit $\om R^2=1/2$, with $C_{\rm{max}}=C_{bh}=1$. This is the endpoint of $C_{\rm{max}}(\om R^2)$, at which it attains its maximum value. Therefore, the allowed range of the maximum compactness associated with the Buchdahl limit is
\beq
\f{4}{9}\leq C_{\rm{max}} \leq 1.
\eeq
It increases monotonically from
$C_{\rm{max}}= 4/9$ in the GR limit $\om R^2 \ra \infty$ to  $C_{\rm{max}}=1$ in the extremal-black-hole limit $\om R^2 \ra 1/2$ \cite{Park:2010,Son:2026,Son:2026b}.

It is interesting to note that (uniform-density) stars exist even for
$0<M <1/\sqrt{2 \om}$ corresponding to
$\om R^2<1/2$, below the extremal black hole limit, $M=1/\sqrt{2 \om}$, $\om R^2=1/2$, which correspond to the minimum black hole mass and radius allowed without a naked singularity.

So far, by requiring the pressure to be finite, {\it positive}, and monotonically {\it non-increasing} with $r$, I have obtained the modified Buchdahl bound on the maximum compactness of the positive-mass stars and $\om > {M}/({2  R^3})$ or $b<0$ that
constraints the allowed static \Ho~stars
with {\it uniform} density and
{\it isotropic} pressure, shown as the region I in Fig. \ref{fig:C_vs_x}; let me call these solutions  {\it Type-I} stars. I now show that there are two additional physical mass/compactness bounds that are genuine to \Ho~gravity.\\

(i) When $\om < {M}/({2  R^3})$ or $b>0$, pressure $p(r)$ is {\it negative}
and monotonically increasing with $r$, satisfying the boundary condition $p(R)=0$.
 The hydrostatic equilibrium is possible because repulsive gravity,
 $a=-N^{-1} N'>0$ is balanced by the attractive pressure force $-p'<0$ in the hydrostatic equation (\ref{P'}) together with $\rho+ p>0$ and $\rho+ 3 p>0$, as shown in Fig. \ref{fig:b>0} (see Appendix {\bf A} for details), so satisfying {\it all} null, weak, strong, and dominant energy conditions.
Since there is no Buchdahl limit in this case,
the static solutions (\ref{P_sol_AdS}) and (\ref{N_sol_AdS}) exist up to the (inner) black hole horizon $r_-$
with {\it no} compactness gap between $C_{\rm{max}}$ and $C_{\rm{bh}}$ for the whole region II
in Fig. 1, in contrast to the region I, where the compactness gap vanishes only in the
extremal limit
 \cite{Son:2026}. The pressure (\ref{P_sol_AdS}) is finite up to the (inner) horizon
$r_-$ but
does not exist beyond the horizon since
it becomes {\it complex-valued}. This gives another physical compactness bound which constrains the allowed
region of solutions for $\om < {M}/({2  R^3})$. This new compactness bound is a genuine effect
of \Ho~gravity because the condition $\om < {M}/({2  R^3})$, or $b>0$, has no GR limit as $\om  \ra \infty$.

Furthermore, for a mass above the extremal black hole mass $M>1/\sqrt{2 \om}$, if the radius is smaller than the inner horizon, $R<r_-$, and $\om < {M}/({2  R^3})$ in
the region II, the ``star" is surrounded by the black hole horizons $r_{\pm}$. Its proper
interpretation for an outside observer is therefore
a {\it regular} (non-singular) black hole solution, sourced by the uniform-density
star at the core region $0<r<r_-$, which is singularity-free, as shown by Eq. (\ref{R}),
while the exterior metric is the usual {\it vacuum} black hole solution.
This is the first example of a regular \Ho~black hole solution without curvature singularities.
The most important difference from the usual construction of regular black holes \cite{Dymn:1992} is that our regular black hole solutions satisfy {\it all} energy conditions: in the conventional regular black holes, the SEC is required to be violated at least to evade Penrose's singularity theorem \cite{Penr:1964}.

This regular black hole solution is {\it ultra-compact}, with compactness $1<C~ <C_{bh}$.
On the other hand, for masses below the extremal black hole mass, $0<M<1/\sqrt{2 \om}$, with $\om < {M}/({2  R^3})$, ultracompact uniform-density stars with $C>1$, as well as $C\leq1$, can exist when $\om R^2 \leq 1/2$, as shown in Fig. \ref{metric_comparson} \footnote{$N$ is $C^1$ but $f$ is $C^0$ due to the discontinuous $\rho$ at $r=R$. If we smooth $\rho$ across $R$, $f$ becomes $C^1$ (see footnote 1).}; let me call these solutions {\it Type-II} stars. Hence, in the region II, I find the {\it Ultra-Compact Objects (UCOs)}, which include Type-II \Ho~stars and regular \Ho~black holes, with $1<C~ (<C_{bh})$ and negative pressure $p<0$, while satisfying all the energy conditions. \\

\begin{figure}
\includegraphics[width=5cm,keepaspectratio]{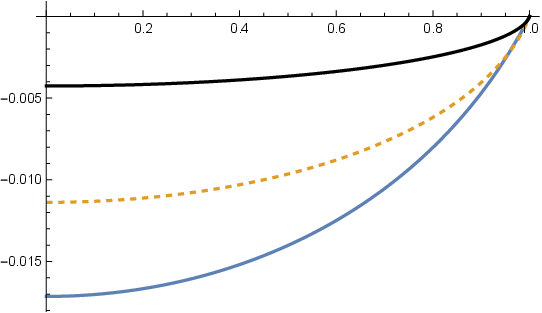}
\includegraphics[width=5cm,keepaspectratio]{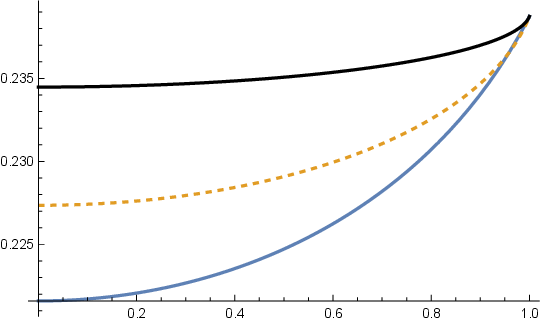}
\includegraphics[width=5cm,keepaspectratio]{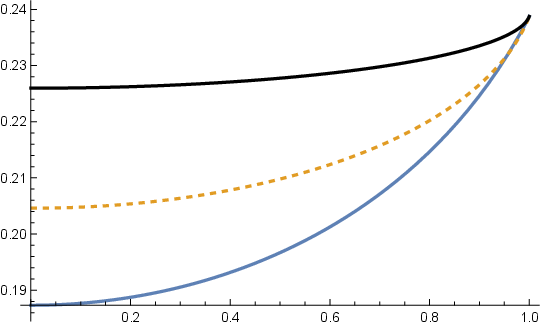}
\caption{The plots of the pressure $p$ (left),
$\rho+p$ (center), and
$\rho+3p$ (right) for $\om < {M}/({2  R^3})$ or $b>0$. The blue solid, yellow dashed, black solid curves represent $\om R^2=1/4,~ 0.9/2,~ 0.99/2$, respectively, with $R=1, M=1$.}
\label{fig:b>0}
\end{figure}

\begin{figure}
\includegraphics[width=7cm,keepaspectratio]{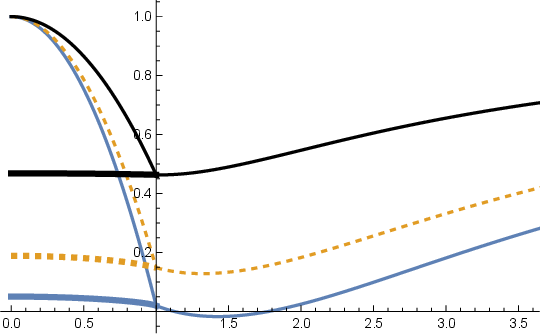}~~~~
\includegraphics[width=7cm,keepaspectratio]{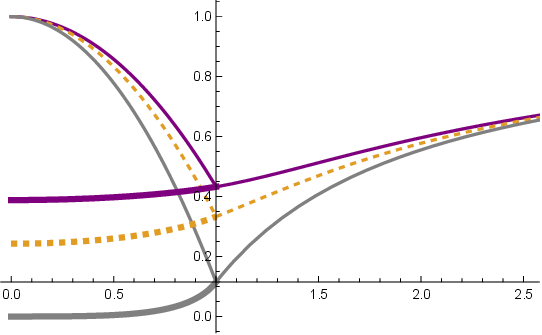}
\caption{The plots of the metric functions $N^2$ (thick lines) and $f$ (thin lines) of the type-II \Ho~stars and regular \Ho~black holes for $\om < {M}/({2  R^3})$ or $b>0$ (the left panel), in comparison with the type-I \Ho~stars for $\om > {M}/({2  R^3})$ or $b<0$ (the right panel). In the left panel, the blue solid, yellow dotted, and black solid curves represent, respectively, the regular black hole and two types of {\it negative-pressure} type-II \Ho~stars with $M/R=1.45,~1.15$ (ultracompact), and $5/9$ for $\om R^2=1/4$ and $R=1$. In the right panel, the purple solid,~ yellow dashed, and gray solid curves represent, respectively, the {\it positive-pressure} type-I \Ho~stars with  $\om R^2=1/2,~1$, and $100$ for $M/R=4/9$ and $R=1$. The type-I stars have the GR limit with $\om R^2\ra
\infty$, whereas the type-II stars and regular \Ho~black hole solutions are genuine to \Ho~gravity and have no GR limit.}
\label{metric_comparson}
\end{figure}

(ii) For a {\it negative} mass $M<0$, the {\it vacuum} black hole solution (\ref{f(R)}) is
 not be acceptable because there is a forbidden region $r< (-4M/\om)^{1/3}$ in which the
solution (\ref{f(R)}) becomes complex-valued, and
there are no black hole horizons to hide this region. However, for a {\it negative-mass star} can still exist if its radius
is larger than the forbidden region, {\it i.e.,} $R>(-4M/\om)^{1/3}$, or equivalently $\om>-4M/R^3$.
In this case, the energy density is negative,
$\rho_0=3M/4 \pi R^3<0$, but the hydrostatic equilibrium is still possible because
$\rho+p<0$,
which violates the null energy condition (NEC), as shown in Fig. 5 (see Appendix {\bf A} for details),
while the repulsive gravity $a=-N^{-1} N'>0$ ensures that the pressure force remains attractive,
$p'=-(\rho+p)N^{-1} N'<0$. In this case, the static solutions (\ref{P_sol_AdS}) and (\ref{N_sol_AdS})
exist up to $R= (-4M/\om)^{1/3}$, as shown by the region III in Fig. 1; beyond this limit, {\it i.e.,}
$R< (-4M/\om)^{1/3}$,
the pressure becomes complex-valued, as in the case beyond the inner black hole horizon for $\om < {M}/({2  R^3})$. Hence, I have another physical bound $R> (-4M/\om)^{1/3}$, or $\om>-4M/R^3$, for $M<0$, as in the regular black hole case for $\om < {M}/({2  R^3}),~M>0$ in the region II.

\begin{figure}
\includegraphics[width=5cm,keepaspectratio]{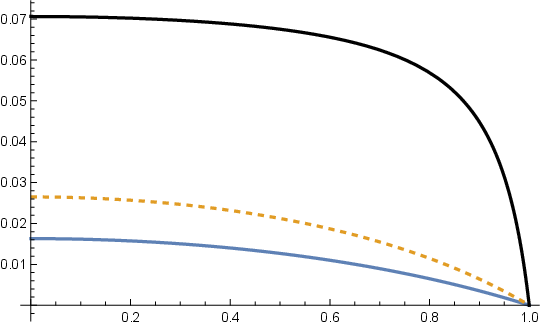}
\includegraphics[width=5cm,keepaspectratio]{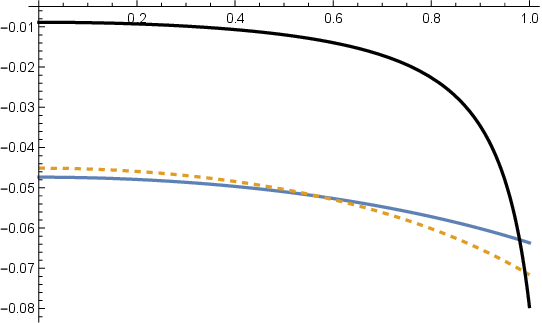}
\includegraphics[width=5cm,keepaspectratio]{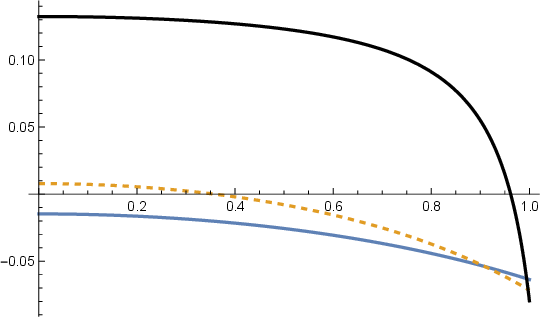}
\includegraphics[width=5cm,keepaspectratio]{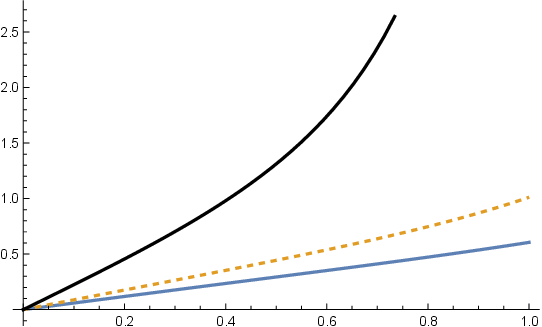}
\includegraphics[width=7cm,keepaspectratio]{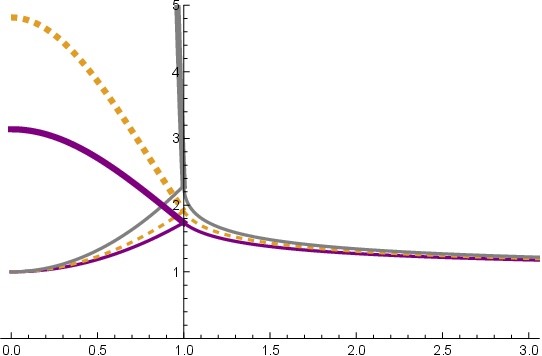}
\caption{The plots of the pressure $p$ (top left),
$\rho+p$ (top center), $\rho+3 p$ (top right),
acceleration $a=-N^{-1} N'$ (bottom left), and the metric functions $N^2$ (thick lines) and
$f$ (thin lines) of \Ho~stars (bottom right)
for $\om>-4M/R^3$ with a {\it negative} mass $M<0$. The blue solid, yellow dashed,
black solid curves represent $M/R=-0.8/3,-0.9/3,0.999/3$, respectively, with $\om R^2=4/3$ and $R=1$.}
\label{fig:M<0}
\end{figure}

\section{
Buchdahl's theorem
}

In GR, one can prove the Buchdahl theorem \cite{Buch:1959}, which states that the maximum
compactness of any reasonable static,
spherically symmetric star with isotropic pressure and monotonically decreasing
density, $d \rho/dr \leq 0$, is given by the Buchdahl bound, $(M/R)_{\rm{max}}=4/9$, attained in
the infinite-central-pressure limit $p(0)=\infty$ for the uniform-density stars.
In this section, I prove that the corresponding theorem
for \Ho~stars whose maximum compactness is given by the {\it modified} Buchdahl bound for uniform-density stars studied in the previous section.

To see whether Buchdahl's theorem can be proved without assuming an equation of state, {\it i.e.},
$p(\rho)$, for the isotropic pressure $p_r=p_{\theta}=p_{\phi} \equiv p(\rho)$, I follow Wald's
approach \cite{Wald:1984} and consider two independent equations that do not {\it explicitly}
involve $p$, either from the full equations of motion (\ref{eom1}), (\ref{eom2}), and (\ref{eom3}), or,
equivalently, from the reduced equations of motion (\ref{Eq_rho}), (\ref{Eq_p}), and (\ref{P'}).
Since Eq. (\ref{eom1}), or equivalently Eq. (\ref{Eq_rho}), has already been solved for $f$ in
Eqs. (\ref{f(r)}) and (\ref{beta_sol}), I now consider the difference between the field equations
in Eq. (\ref{eom3}), so that
the $p$ 
terms from $T_{ij}$ on the left-hand sides cancel and do not appear explicitly:
\beq
0 &=& \f{1}{N} \left(f E_{rr}- \f{1}{r^2} E_{\theta \theta}\right) \no \\
 &=& -\f{\kappa^2 \mu^2}{32 r^4 N} \left\{ (f-1+\om r^2) [2 (1-f)+r f' ] N  \right. \no \\
 && \left. +r \left[\left(2 f (-3-\om r^2 +3 f)+r (1+ \om r^2-3f) f' \right) N' +2 r (1+\om r^2 -f) f N''\right]
 \right\}.
 \label{Eq_0}
\eeq
Introducing
${\cal H}$ and ${\cal W}$ as,
\beq
{\cal H}(r)& \equiv &\f{1}{r^2} \left(f-1-\f{r}{2} f' \right), \label{H}\\
{\cal W}(r) &\equiv & \f{1}{r} \left[f^{1/2} N''+ \left(\f{1}{2} f^{-1/2} f'-r^{-1} f^{1/2}\right) N' \right] \no \\
&=& \left[\f{f^{1/2}}{r} N' \right]', \label{W}
\eeq
Eq. (\ref{Eq_0}) can be written as
\beq
0 =\f{\kappa^2 \mu^2 \om}{16}
\left\{
\left[ 1+ \f{f-1}{\om r^2} \left(1-2 r f N^{-1} N' \right)\right] {\cal H}
+r f^{1/2}N^{-1} \left(-1 +\f{f-1}{\om r^2}\right) {\cal W}
\right\}
\eeq
and we then obtain the master equation
\beq
{\cal W}= \left(
\f{1+ \f{f-1}{\om r^2} \left(1-2 r f N^{-1} N'\right) }{r f^{1/2}N^{-1} \left( 1 -\f{f-1}{\om r^2}\right)}
\right) {\cal H},
\label{W=AH}
\eeq
or
\beq
\left[\f{f^{1/2}}{r} N' \right]'=\left(
\f{1+ \f{f-1}{\om r^2} \left(1-2 r f N^{-1} N'\right) }{r f^{1/2}N^{-1} \left( 1 -\f{f-1}{\om r^2}\right)}\right)
\left[ \f{r}{4 \sqrt{\om^2 + \f{\be}{ r^{3}} }} \left( \f{\be}{ r^3}\right)'\right]
\label{W=AH2}
\eeq
by substituting the solution (\ref{f(r)}) for $f$ in ${\cal H}$. It is rather remarkable that
the functions ${\cal H}$ and ${\cal W}$ of Eqs. (\ref{H}) and
(\ref{W}), respectively, satisfy almost the same equation (\ref{W=AH2}) in GR \cite{Buch:1959}, relating $(r^{-1} f^{1/2} N' )'$ to a radial derivative of average mass density $\propto (\be/r^3)'$, which can be interpreted as a radial conservation equation.

Integrating Eq. (\ref{W=AH2}) inward from the star's surface $R$ to an interior
radius $r$, I obtain
\beq
\f{f^{1/2} (r)}{r} N'(r)-\f{f^{1/2} (R)}{R} N'(R)= \int^r_R \chi (r) \left( \f{m(r)}{r^3}\right)',
\label{master_eq}
\eeq
where
\beq
\chi(r)  \equiv \f{1+ \f{f-1}{\om r^2} \left(1-2 r f N^{-1} N'\right) }{r f^{1/2}N^{-1} \left( 1 -\f{f-1}{\om r^2}\right)} \f{r}{\sqrt{1+ \f{4 m(r)}{\om r^{3}} }}
\label{weight}
\eeq
and $\be(r)\equiv 4 \om m(r)$, with $\om>0$, has been used.

Let us {\it suppose} that the right-hand side of Eq. (\ref{master_eq}) is {\it non-negative}, then I have
\beq
\f{f^{1/2} (r)}{r} N'(r) \geq \f{f^{1/2} (R)}{R} N'(R)\equiv {\cal M},
\label{Buch_eq}
\eeq
where
\beq
{\cal M} [M,R, \om]= \om \left[ \f{\sqrt{1+\f{4M}{\om R^3}}-1-\f{M}{\om R^3}}{\sqrt{1+\f{4M}{\om R^3}}}\right]
\label{M}
\eeq
is computed from the exterior black hole metric (\ref{f(R)}), using the continuity of the metric and its derivative at $r=R$. Multiplying $r f^{-1/2}$ to Eq. (\ref{Buch_eq}) and integrating inward again from $R$ to $0$, I obtain
\beq
N(0) \leq N(R)-{\cal M} \int^R_0 \left[1+\om r^2 \left(1-\sqrt{1+ \f{4 m(r)}{\om r^{3}} } \right) \right]^{-1/2} r dr
\label{N0vsNR}
\eeq
for any $f(r)=1+\om r^2 \left(1-\sqrt{1+ {4 m(r)}/{\om r^{3}} } \right)>0$ that satisfies the staticity condition and for arbitrary ${\cal M}$; here, I have also used solution (\ref{f(r)}) for $f$.

Let me now assume that the average density, which is proportional to $m(r)/r^3$, is {\it non-increasing} with $r$, {\it i.e.}, $(m(r)/r^3)' \leq 0$, so that
\beq
\f{m(r)}{r^3} \geq \f{M}{R^3}
\eeq
by integrating inward from $R$ to an interior radius $r$. Then, for $M/R<2 \om R^2$, or equivalently $b<0$ in the uniform density star solution, so that ${\cal M}>0$, the second term on the right-hand side of Eq. (\ref{N0vsNR}) reaches its maximum (or the minimum magnitude) when $m(r)/r^3=M/R^3$. Hence, I obtain
\beq
N(0) &\leq& N(R)-{\cal M} \int^R_0 \left[1+\om r^2 \left(1-\sqrt{1+ \f{4 M}{\om R^{3}} } \right) \right]^{-1/2} r dr \label{N0vsNR2} \no \\
&\leq& N(R)-\f{{\cal M}}{\om \left(1-\sqrt{1+ \f{4 M}{\om R^{3}} } \right) }
\left[ \sqrt{1+\om R^2 \left(1-\sqrt{1+ \f{4 M}{\om R^{3}} } \right) }-1\right]  \no \\
&\leq& \left\{1+ \left[ \f{\sqrt{1+ \f{4 M}{\om R^{3}} } -1- \f{M}{\om R^{3}}  }{\sqrt{1+ \f{4 M}{\om R^{3}} } \left(\sqrt{1+ \f{4 M}{\om R^{3}} } -1 \right)} \right] \right\}N(R)-\left[ \f{\sqrt{1+ \f{4 M}{\om R^{3}} } -1- \f{M}{\om R^{3}}  }{\sqrt{1+ \f{4 M}{\om R^{3}} } \left(\sqrt{1+ \f{4 M}{\om R^{3}} } -1 \right)} \right],
\eeq
where I have used $N(R)=\sqrt{1+\om R^2 \left(1-\sqrt{1+ \f{4 M}{\om R^{3}} } \right) }$ from the exterior black hole metric. To satisfy the {\it staticity} condition $N^2>0$, $N$ must also be {\it positive} also, because it could be negative only by passing through $N=0$, which contradicts $N^2>0$. Thus, from $N(0) \geq 0$, the necessary condition for staticity is
\beq
&&\left\{1+ \left[ \f{\sqrt{1+ \f{4 M}{\om R^{3}} } -1- \f{M}{\om R^{3}}  }{\sqrt{1+ \f{4 M}{\om R^{3}} } \left(\sqrt{1+ \f{4 M}{\om R^{3}} } -1 \right)} \right] \right\}\sqrt{1+\om R^2 \left(1-\sqrt{1+ \f{4 M}{\om R^{3}} } \right) }  \no \\
&&\geq \left[ \f{\sqrt{1+ \f{4 M}{\om R^{3}} } -1- \f{M}{\om R^{3}}  }{\sqrt{1+ \f{4 M}{\om R^{3}} } \left(\sqrt{1+ \f{4 M}{\om R^{3}} } -1 \right)} \right],
\eeq
which becomes the inequality
\beq
\left(\f{3M}{\om R^3} \right) \sqrt{1+\om R^2 \left(1-\sqrt{1+ \f{4 M}{\om R^{3}} } \right) }
\geq \sqrt{1+ \f{4 M}{\om R^{3}} } -1- \f{M}{\om R^{3}}.
\eeq
The equality gives the maximum compactness of $M_{\rm{max}} (\om,R)/R$, given by the modified Buchdahl bound in Eq. (\ref{Buchdahl_limit_Q}), attained in the infinite-central-pressure limit $p(0)=\infty$ for
uniform-density stars; this proves
Buchdahl's theorem.

In Eq. (\ref{master_eq}), I have assumed a weighted
average-density condition
\beq
-\int^R_r \chi(r) \bar{\rho}'(r) \geq 0
\label{averaged_condition}
\eeq
with a weight factor $\chi$ for the average density $\bar{\rho}
\propto {m}/{r^3}$ and $0<M/R<2 \om R^2$, in addition to  the usual conditions that the average density is {\it non-increasing}, $\bar{\rho}'\leq 0$, and that $N>0$ from the staticity condition. Because the weight factor $\chi$ in Eq. (\ref{weight}) depends on $N'$, our proof, unlike Buchdahl's theorem in GR \cite{Buch:1959}, is {\it not} completely independent of the equation of sate $p(\rho)$ entering through Eq. (\ref{Eq_p}), {\it i.e.}, \beq
\chi(r) \propto \left[ 1+\left(1-\sqrt{1+ \f{4 m(r)}{\om r^{3}}}\right)
\left(1+2r f \f{ p'(r)}{p(r)+\rho(r)} \right) \right],
\eeq
up to
{\it positive}-definite multiplicative factors for $\rho>0$ or $m>0$. If ${m}/{r^3} \leq (3/4)\om$ holds together with the usual non-increasing pressure, $p'\leq 0$, and the NEC, $p+\rho>0$, then $\chi\geq 0$ can be found, and hence condition (\ref{averaged_condition}) is automatically satisfied. At the star's surface $r=R$, however, it implies $M/R\leq(3/4)\om R^2$, which is more restrictive than the condition $M/R<2\om R^2$ that we have assumed in our proof. Thus, by assuming the condition (\ref{averaged_condition}), I require an additional {\it weighted-monotonicity} condition on the average density: if \(\chi\) becomes negative in some region,
the density gradient $\bar{\rho}'$ must be sufficiently suppressed there or
compensated by regions where \(\chi>0\) or by positive pressure gradient, $p'/(p+\rho)>0$.

\section{Concluding Remarks}

I have obtained the interior solution for incompressible, spherically symmetric static stars
in four-dimensional $z=3$ non-projectable \Ho~gravity with an arbitrary cosmological constant, isotropic pressure, and $\la=1$, in which  Birkhoff's
theorem holds. For a vanishing cosmological constant, I have found several new phases
of \Ho~stars and black holes that are genuine to \Ho~gravity. These include {\it ultra-compact objects (UCOs)} of the Type-II stars and regular black holes with
compactness $1<C~(< C_{\rm{bh}})$ and negative pressure $p<0$ while, surprisingly, satisfying
all (null, weak, strong, and dominant) four standard energy conditions. I have also found {\it negative-mass}
Type-III stars and the usual Type-I stars with positive pressure, where the modified Buchdahl bound of the maximum compactness is
$4/9 \leq C_{\rm max} \leq 1$ for $p>0$
\cite{Park:2010,Son:2026}.
The {\it regular} \Ho~black-hole solution is the long-sought
non-singular black hole solution: its interior metric has no curvature singularities
but does not require exotic matter that violates the energy conditions, while its exterior
metric is
identical to that of the corresponding {\it vacuum} \Ho~black-hole solution.

I have obtained the phase diagram for incompressible (uniform density) stars.
Although the quantitative details may depend on the equation of
state, I expect the qualitative phase structure found here to persist
for more general stellar matter.
Extensions to asymptotically {\it (A)dS} spacetimes would be interesting, since the
corresponding phase structures are richer but more involved. An analysis of the dynamical stability
of \Ho~stars and regular \Ho~black holes will also be important
for their viability in nature.
In particular, if the regular \Ho~black-hole solutions are stable, it would be
interesting
to investigate their role as primordial black holes.

I have proved Buchdahl's theorem for a more general class of static, spherically symmetric, and isotropic
Type-I stars
without assuming an explicit form of the equation of state. The proof uses the weight-monotonicity condition on the average density, in addition to the usual non-increasing average density. It would be challenging to find a simpler derivation that does not reply on the weight-monotonicity condition, if such a derivation exists, or to investigate whether the condition is valid for the more realistic equations of state \footnote{A preliminary computation shows that, for four representative equations of state--AP4, WFF1, MPA1, and SLy4--the weight-monotonicity condition is satisfied at the fixed central density $\rho(0)=6.0 \times10^{14}~ {\rm g~ cm^{-3}}$ for values of $\om$ above the minimum value $\om_{\rm min} \cong 1.22 \times 10^{-3}~ {\rm km^{-2}}$. The value of $\om_{\rm min}$ suggests that the condition is satisfied in the numerical analysis of Refs. \cite{Son:2026, Son:2026b}, which use $\om=(1/32)~ {\rm km^{-2}}$ and $\om=(1/0.32)~{\rm km^{-2}}$. However, to establish that the computed threshold is the true global minimum, the central density must be varied in the full {\it M-R} families for each equation of state at every tested value of $\om$. } \cite{Son:2026b}
\\

{\it Note added}: After completing the original version of this work in 2010 \cite{Park:2010},
a related paper has recently appeared in which the uniform-density star
solution for a vanishing cosmological constant was also obtained
\cite{Son:2026}.
Motivated by that work, I revisited my 2010 analysis and identified an
error in
the Buchdahl bound near the extremal-black-hole limit. The present version corrects this error and provides a
clearer understanding of the modified Buchdahl bound, together with an
{\it analytic} proof of Buchdahl's theorem in \Ho~gravity, which has been investigated only
{\it numerically} in Refs. \cite{Son:2026, Son:2026b}.

\section*{Acknowledgments}

I would like to thank Edwin J. Son for helpful correspondences regarding his work.
This work was supported by the Basic Science Research Program through the National
Research Foundation of Korea (NRF) funded by the Ministry of Education,
Science and Technology (NRF-2020R1A2C1010372, 2020R1A6A1A03047877).

\appendix

\section{Details on Energy Conditions}
\label{app:Energy-condition}

In this appendix, I present details of the energy conditions studied in the text. To that end, I first note that the pressure for an incompressible-star solution with a vanishing cosmological constant is given by
\beq
p(r)=\f{3M}{4 \pi R^3} \f{B-A}{C+A},
\label{p_flat}
\eeq
where $A=\sqrt{1+Q r^2}, B=\sqrt{1+Q R^2}$, and
\bea
C=\f{\f{12 M}{\om R^3} \sqrt{1+Q R^2}}{\left(1-\sqrt{1+\f{4M}{\om R^3}}\right) \left(3-\sqrt{1+\f{4M}{\om R^3}}\right)}
\eea
with $Q=\om \left(1-\sqrt{1+\f{4M}{\om R^3}}\right)$.

For $M>0$, I have $Q<0$ (note $\om>0$) and hence $B-A\leq0$ since $0<r\leq R$.
In this case, if $M/\om R^2>2$ is also considered, I find $C>0$ and hence $C+A>0$,
so that the pressure
(\ref{p_flat}) is finite and non-positive $p(r)\leq 0$ ($p(r)=0$ only for $r=R$,
regardless of $M$). Whereas, if $(0<) M/\om R^2<2$ is considered, I find $C<0$
and hence $C+A \leq 0$ is also possible, so that $p$ can be positive or even
infinite, {\it i.e.}, $0<p \leq \infty$. If $r_*$ is defined by $C+A(r_*)=0$, I find $C+A<0$ for $r<r_*$ and $C+A>0$ for $r<r_*$, such that
$p(r_*)=+\infty$ and $p(r)>0$ for $r>r_*$, whereas $p(r)<0$ for $r<r_*$.
The Buchdahl limit (\ref{y_true}) corresponds to $r_*=0$, {\it i.e.}, infinite pressure at the center.

For $M<0$, I have $Q>0$ and hence $B-A>0$. In this case, I find $C<0$, but $C+A<0$ is
allowed only because $M/\om R^3>-1/4$ is required for $C$ to be real so that $p>0$.

Let me now define the energy-condition function
\beq
EC_{\eta} (r) &\equiv& \rho+ \eta p(r) \no \\
&=&\f{3M}{4 \pi R^3} \f{(C+\eta B)-(\eta-1)A}{C+A},
\eeq
where
\beq
C+\eta B=\f{4 \eta \left(1-\sqrt{1+\f{4M}{\om R^3}}+(1+3 \eta^{-1})\f{M}{\om R^3}\right) }{\left(1-\sqrt{1+\f{4M}{\om R^3}}\right) \left(3-\sqrt{1+\f{4M}{\om R^3}}\right)}.
\eeq

For $\eta=1$, I have
\beq
EC_1(r)&\equiv& \rho+ p(r) \no \\
&=&\f{3M}{4 \pi R^3} \f{C+B}{C+A},
\eeq
where
\beq
C+B=\f{4 \left(1+\f{4M}{\om R^3}-\sqrt{1+\f{4M}{\om R^3}}\right) }{\left(1-\sqrt{1+\f{4M}{\om R^3}}\right) \left(3-\sqrt{1+\f{4M}{\om R^3}}\right)}.
\eeq
Since the numerator of $C+B$ is positive for $M>0$, I find  $C+B<0$ for $0<M/\om R^3<2$ and $C+B>0$ for $M/\om R^3>2$, so that $EC_1>0$ $(r>r_*)$ or $EC_1<0$ $(r<r_*)$ for $0<M/\om R^3<2$, whereas  $EC_1>0$ always for $M/\om R^3>2$. Hence, the null energy condition (NEC), $EC_1>0$, is satisfied for $r>r_*$ in both cases, $0<M/\om R^3<2$ and $M/\om R^3>2$.
On the other hand, for $M<0$, I find $C+B<0$ and hence $EC_1<0$, so that NEC is
violated.

The satisfaction of NEC for $0<M/\om R^3<2$ $(r>r_*)$ or its violation for $M<0$ is not surprising since $\rho, p>0$ (trivial case) or $\rho<0, p>0$, respectively. However, the satisfaction of NEC for $M/\om R^3>2$ is rather surprising since it implies the dominant energy condition (DEC) as well, which corresponds to the relativistic causality bound or speed limit, because $EC_{-1}=\rho-p>0$ is trivially satisfied due to $\rho>0, p<0$.\\

For $\eta=-1$, we have
\beq
EC_{-1}(r)&\equiv& \rho- p(r) \no \\
&=&\f{3M}{4 \pi R^3} \f{C-B+2A}{C+A},
\eeq
where
\beq
C-B=\f{-4 \left(1-\f{2M}{\om R^3}-\sqrt{1+\f{4M}{\om R^3}}\right) }{\left(1-\sqrt{1+\f{4M}{\om R^3}}\right) \left(3-\sqrt{1+\f{4M}{\om R^3}}\right)}.
\eeq
The numerator of $C-B$ is negative for $M>0$ and hence $C-B<0$ for either $0<M/\om R^3<2$ or $-1/4<M/\om R^3<0$, while $C-B>0$ for $M/\om R^3>2$. Hence, for $M/\om R^3>2$, it is easy to see $EC_{-1}>0$, as well as $EC_1>0$, {\it i.e.}, DEC is satisfied because $C+A>0$. On the other hand, for $-1/4<M/\om R^3<0$, I find $C-B+2A \leq C+B<0$, due to $A\leq B$ for all $0 \leq r<R$, so that $EC_{-1}<0$, as well as $EC_{1}<0$; hence, DEC is completely violated.

For $0<M/\om R^3<2$, I first note that $EC_{-1}(R)=\f{3M}{\om R^3}>0$ at $r=R$. However, $EC_{-1}$ can be {\it negative} in an intermediate region $r>r_*$, where $C+A(r)<0$ while $C-B+2A>0$, with a pole $r_*$ satisfying $C+A(r_*)=0$. Let $r_*^+=r_*+\ep~ (0<\ep \ll 1)$ denotes a point just outside $r_*$. Then, $C+A \sim -\ep<0$ and hence $EC_{-1}\sim \f{3M}{\om R^3} (\f{-\ep+A-B}{\ep})>0$ from $A-B>0$, so that $EC_{-1} (r_*^+)\ra +\infty$ near $r=0$. Therefore, because $EC_{-1}>0$ and $EC_1>0$ for $0<M/\om R^3<2$ $(r>r_*)$, DEC is satisfied near $R$ but is partially violated in the interior region near $r_*$, where $EC_{-1}<0$. Actually, this is a common feature of uniform-density stars with the pole $r_*$ of $p(r_*)=\infty$ because $p(r)$ grows from $p(R)=0$ toward  $p(r_*)=\infty$, so that $EC_{-1}=\rho-p(r)$ decreases from $EC_{-1}=\rho>0$ until $EC_{-1}=0$ with $\rho=p(r)>0$ at $r=r_c$ and $EC_{-1}<0$ for $0<~ \rho<p(r)$ at $r<r_c$. \footnote{The critical radius $r_c$ where $EC_{-1}$ changes its sign obtained by solving $C-B+2A=0$ and is given by $r_c=\sqrt{(B-C)^2-4}/4Q$ .}

For $\eta=3$, I have
\beq
EC_{3}(r)&\equiv& \rho + 3 p(r) \no \\
&=&\f{3M}{4 \pi R^3} \f{C+3B-2A}{C+A}.
\eeq
Since $C+3B-2A=C+B+2(B-A)>0$ for $M/\om R^3>2$ and $C+B+2(B-A)<0$ for $-1/4<M/\om R^3<0$, I find $EC_3>0$ and $EC_3<0$, respectively. On the other hand, from $C+3B-2A<0$ for $0<M/\om R^3<2$, I find $EC_3>0$ $(r>r_*)$ and $EC_3<0$ $(r<r_*)$. Hence, SEC is satisfied only for $M/\om R^3>2$ or  $0<M/\om R^3<2$ $(r>r_*)$.
All results are summarized in Table \ref{tab:energy condition summary}.

\begin{table}[t]
\begin{tabular}{c| c| c| c| c| c| c| c| c}
\hline
$M/\om R^3$ & $EC_1$ &  $EC_{-1}$ &  $EC_{3}$ &  $p$ & \makecell{NEC \\$\rho+p \geq 0$} & \makecell{WEC \\$\rho \geq 0, \rho+p \geq 0$} & \makecell{DEC \\ $\rho \geq |p|$} &
\makecell{SEC\\ $\rho+p \geq 0, \rho+3p \geq 0$} \\
\hline\hline
$M/\om R^3>2$ &+  &+ &+ &-- &Y &Y &Y &Y\\
\hline
$ \makecell{M/\om R^3<2 \\(r>r_*)}$  &+  &--/+ &+ &+ &Y &Y &N/Y &Y \\
\hline
$\makecell{M/\om R^3<2 \\(r<r_*)}$  &+  &--/+ &-- &-- &N &N &N &N\\
\hline
$\makecell{-1/4<M/\om R^3<0}$ &--  &-- &-- &+ & N &N &N &N  \\
\hline
\end{tabular}
\caption{Summary of the energy-condition functions $EC_1$, $EC_{-1}$, $EC_{3}$, and all energy conditions.}
\label{tab:energy condition summary}
\end{table}

\newcommand{\J}[4]{#1 {\bf #2} #3 (#4)}
\newcommand{\andJ}[3]{{\bf #1} (#2) #3}
\newcommand{\AP}{Ann. Phys. (N.Y.)}
\newcommand{\MPL}{Mod. Phys. Lett.}
\newcommand{\NP}{Nucl. Phys.}
\newcommand{\PL}{Phys. Lett.}
\newcommand{\PR}{Phys. Rev. D}
\newcommand{\PRL}{Phys. Rev. Lett.}
\newcommand{\PTP}{Prog. Theor. Phys.}
\newcommand{\hep}[1]{ hep-th/{#1}}
\newcommand{\hepp}[1]{ hep-ph/{#1}}
\newcommand{\hepg}[1]{ gr-qc/{#1}}
\newcommand{\bi}{ \bibitem}

\end{document}